\documentclass[letter,twocolumn]{jpsj2} %% two-column layout
\title{X-ray Irradiation-induced Carrier Doping Effects in Organic Dimer-Mott Insulators}

\author{Takahiko \textsc{Sasaki}$^{1}$\thanks{E-mail address: takahiko@imr.tohoku.ac.jp}, Hajime \textsc{Oizumi}$^{1}$, Naoki \textsc{Yoneyama}$^{1}$, Norio \textsc{Kobayashi}$^{1}$ and Naoki \textsc{Toyota}$^{2}$}

\inst{$^{1}$Institute for Materials Research, Tohoku University, Katahira 2-1-1, Aoba-ku, Sendai 980-8577\\

$^{2}$Physics Department, Graduate School of Science, Tohoku University, Aoba-ku, Sendai 980-8578}

\abst{We report X-ray irradiation-induced carrier doping effects on the electrical conductivity in the organic dimer-Mott insulators $\kappa$-(ET)$_{2}$$X$ with $X =$ Cu[N(CN)$_{2}$]Cl and Cu$_{2}$(CN)$_{3}$.
For $\kappa$-(ET)$_{2}$Cu[N(CN)$_{2}$]Cl, we have observed a large decrease of the resistivity by 40 \% with the irradiation at 300 K and the metal-like temperature dependence down to about 50 K. 
The irradiation-induced defects expected at the donor molecule sites might cause a local imbalance of the charge transfer in the crystal.
Such molecular defects result in the effective doping of carriers into the half-filled dimer-Mott insulators.  
}

\kword{Mott insulator, organic conductor, BEDT-TTF, carrier doping, X-ray irradiation}

\begin{document}
\maketitle

%Introduction
A metal - insulator (MI) transition has been one of the central issues in strongly correlated electron system.
Among the various types of the MI transitions, the Mott transition has been the most attractive phenomenon arising from the electron-electron interactions in a wide range of materials.\cite{Imada}
A Mott insulator derives from the large Coulomb repulsion energy $U$ in comparison with the band width $W$. 
Two different types of the Mott transitions exist: one can change the strength of the interaction $U/W$ with keeping the carrier-filling its commensurate value, referred to as a band-width controlled Mott transition, and one can introduce carriers to the commensurate density with keeping the strength of the interaction constant, referred to as a band-filling controlled one.  
The former case, for example, can be seen in vanadium oxides and molecular conductors and the latter in high-$T_{\rm c}$ copper oxides.  

Organic charge transfer salts based on the donor molecule BEDT-TTF, bis(ethylenedithio)-tetrathiafulvalene abbreviated ET, have been recognized as one of the highly correlated electron system.\cite{Miyagawa} 
Among them, $\kappa$-(ET)$_{2}$$X$ with $X =$ Cu(NCS)$_{2}$, Cu[N(CN)$_{2}$]$Y$ ($Y =$ Br and Cl), Cu$_{2}$(CN)$_{3}$, etc. has attracted considerable attention from the point of view of a strongly correlated quasi-two-dimensional electron system because the strong dimer structure consisting of two ET molecules makes the conduction band close to half-filled with the effective Coulomb energy $U_{\rm eff}$ on the dimer.\cite{Miyagawa,Kanoda2,Kino}  
Furthermore softness of the lattice with high compressibility enables us to modulate the band width easily by small physical and chemical pressures with keeping the band-filling unchanged.\cite{Ito,Lefebvre,Yoneyama1}  
Thus the $\kappa$-(ET)$_{2}$$X$ system has been considered to be one of the typical band-width controlled Mott systems.  

The modulation of the band-filling has been actively examined for the molecular materials because the carrier doping into the insulators and also into the conductors must be important to tune the electronic functions. 
A carrier injection by the field effect technique has been extensively studied in the organic crystals for aiming at the future electronic devices.\cite{FET}
Another way to control the band-filling particularly in the charge transfer salt is the chemical synthesis for making the salt with nonstoichiometric composition.\cite{Mori}  
But this chemical approach usually cannot control the composition arbitrarily.  

The irradiation can modify the crystalline properties of molecular materials by the electronic excitation and/or collisions with accelerated ions.\cite{x-ray}  
The irradiation effects have been studied for the organic materials from both viewpoints of the molecular defects and its relation to carrier scatterings.\cite{TMTSF,BEDT}  
Recently from X-ray irradiation studies for the organic superconductor $\kappa$-(ET)$_{2}$Cu(NCS)$_{2}$, Analytis {\it et al.} claimed the unconventional superconductivity with the large reduction of $T_{\rm c}$ by the irradiation-induced non-magnetic defects.\cite{Analytis}  
More interestingly their experiments showed a substantial decrease of the resistivity in the high temperature region.  
As they claimed a violation of the Matthiessen's rule at high temperatures, this observation contradicted a naive idea for the transport properties in the case of increasing the carrier scattering by the defects. 
The similar anomalous resistive behavior in the superconducting $\kappa$-(ET)$_{2}$$X$ at high temperatures has been discussed with the relation to the sample quality.\cite{Strack}  

In this letter we report X-ray irradiation effects on the transport properties in the organic dimer-Mott insulators $\kappa$-(ET)$_{2}$$X$.
The expected defects at donor molecule sites might cause a local imbalance of the charge transfer between ET donor and anion $X$ molecules.  
Such molecular defects result in the effective carrier doping into the half-filled dimer-Mott insulators.

%Experiments

Single crystals of $\kappa$-(BETS)$_{2}$FeCl$_{4}$, $\kappa$-(ET)$_{2}$$X$ with $X =$ Cu(NCS)$_{2}$, Cu[N(CN)$_{2}$]Cl and Cu$_{2}$(CN)$_{3}$ were grown by the standard electrochemical oxidation method, where BETS is  bis(etylenedithio)-tetraselenafulvalene.
The in- and inter-plane resistivities were measured by a dc four-terminal method.
Samples were irradiated at 300 K by using a non-filtered copper or tungsten target with 40 kV and 20 mA.  
The dose rate in the present experiments was expected to be about 0.5 MGy/hour following the comparison to the previous report by Analytis {\it et al.}\cite{Analytis}
During the irradiation, the resistivity was monitored as the irradiation time dependence.  
After each step (5 - 7 days) of the irradiation, the temperature dependence of the resistivity was measured at the same cooling rate of $- 0.2$ K/min in order to avoid a structural disorder effect due to the fast cooling.\cite{Mueller}

%Results and Discussions

\begin{figure}[tb]
\begin{center}
\includegraphics[viewport=3cm 12.5cm 17cm 22cm,clip,width=1.0\linewidth]{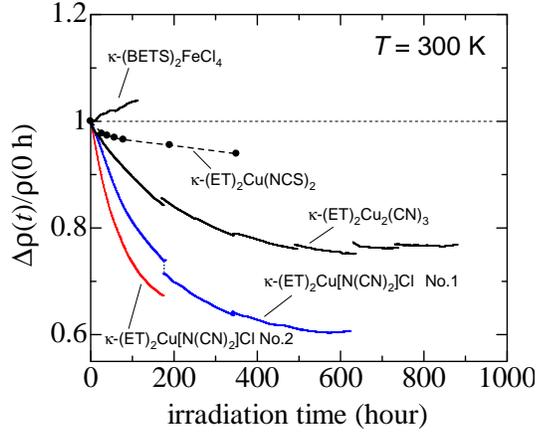}
\end{center}
\caption{(Color online) The X-ray irradiation time dependence of the in-plane resistivity at 300 K. Each resistivity curve is normalized by the value before the irradiation. Copper and tungsten targets (40 kV, 20mA) are used for $\kappa$-(ET)$_{2}$Cu$_{2}$(CN)$_{3}$, $\kappa$-(ET)$_{2}$Cu[N(CN)$_{2}$]Cl (No. 2) and $\kappa$-(BETS)$_{2}$FeCl$_{4}$, and for $\kappa$-(ET)$_{2}$Cu(NCS)$_{2}$ and $\kappa$-(ET)$_{2}$Cu[N(CN)$_{2}$]Cl (No.1), respectively}
\label{fig1}
\end{figure}

Figure 1 shows the irradiation time dependence of the in-plane resistivity of the $\kappa$-type ET and BETS salts.  
In all samples measured, the appearance of the sample surface and the resistance of the electrical contacts did not change with the irradiation.
We have checked the reproducibility of the results reported for the superconducting $\kappa$-(ET)$_{2}$Cu(NCS)$_{2}$.\cite{Analytis}  
We have quantitatively reproduced well their observations in the interlayer resistivity; 1) the decrease of $T_{\rm c}$, 2) the increase of the residual resistivity, 3) the reduction of the magnitude of the resistivity broad peak around 100 K, and 4) the decrease of the resistivity at high temperatures.  

The results on the superconductivity by Analytis {\it et al.}\cite{Analytis} have been explained on the basis of the pair breaking effect by non-magnetic impurity (defect) scatterings in the non-$s$-wave superconductivity.\cite{Powell2}  
The irradiation defects, which serve as an additional source of electron scatterings to enhance the resistivity, can be confirmed also by the results in the organic conductor $\kappa$-(BETS)$_{2}$FeCl$_{4}$.  
This salt has a smaller ratio $U_{\rm eff}/W$ in comparison to the $\kappa$-type BEDT salts and then it is situated far from the Mott transition.  
The resistivity shows a simple metallic behavior from room temperature down to 2 K at least.\cite{BETS}  
After the irradiation for 110 hours, the resistivity increases by 4 \% at 300 K as shown in Fig. 1, and the residual resistivity does by 26 \%, which is obtained by an extrapolation to 0 K by following an observed $T^{2}$-dependence below 20 K (the results are not shown here). 
Similar increase of the resistivity due to the defects induced by the irradiation has been observed in many other organic conductors.\cite{x-ray,BEDT}

Analytis {\it et al.}\cite{Analytis} proposed a defect-assisted interlayer conduction channel model\cite{x-ray} for the reduction of the resistivity at high temperatures. 
This model expects a change of the anisotropy between in- and inter-plane resistivity.  
We found, however, in the present study that the anisotropy of about 100 at 300 K does not change so much with increasing the radiation dose.  
This observation is not consistent with this model.  
We would propose here that the decrease of the resistivity in $\kappa$-(ET)$_{2}$Cu(NCS)$_{2}$ has the same origin with that in the insulators as follows although the magnitudes are different.  

The Mott insulators of $\kappa$-(ET)$_{2}$$X$ with $X =$ Cu[N(CN)$_{2}$]Cl (ref. \citen{Williams}) and Cu$_{2}$(CN)$_{3}$ (ref. \citen{Bu}) show a remarkable reduction of the resistivity at 300 K in comparison with that in $\kappa$-(ET)$_{2}$Cu(NCS)$_{2}$.  
In both insulators, the resistivity decreases rapidly in the initial irradiation, and then tends to become saturated.
After several hundred hours irradiation, the resistivity takes a minimum and turns to increase slightly. 
These features are qualitatively same in both cases using the copper and tungsten targets although the former contains characteristic $K_{\alpha}$ radiation and the latter does not.
Some differences of the resistivity are found in between steps of the irradiation.  
Though the reason is not clear at present, it may be an aging effect by the partial restoration of the irradiation damages.

\begin{figure}[tb]
\begin{center}
\includegraphics[viewport=4cm 5cm 18cm 24cm,clip,width=0.9\linewidth]{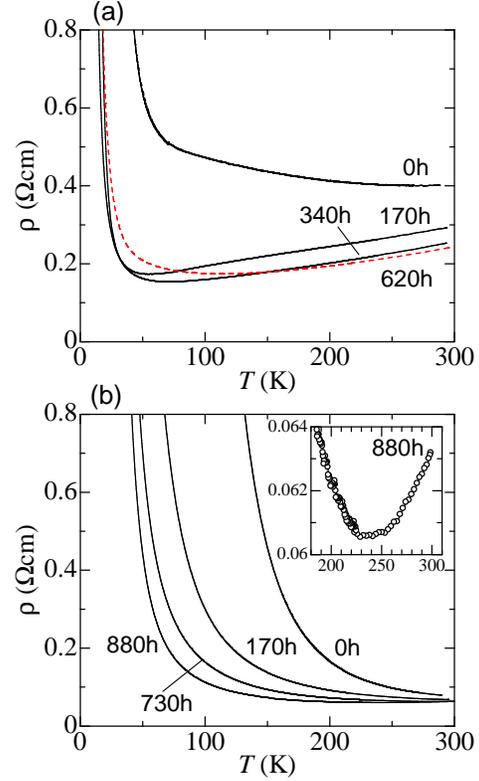}
\end{center}
\caption{(Color online) Temperature dependence of the in-plane resistivity in (a) $\kappa$-(ET)$_{2}$Cu[N(CN)$_{2}$]Cl (No.1) and (b) $\kappa$-(ET)$_{2}$Cu$_{2}$(CN)$_{3}$.  The time indicated at each curve is the total irradiation time.  The broken curve in Fig. 2(a) corresponds to the data after total 620 hour irradiation.  The inset in Fig.2 (b) is the enlarged view of the data in high temperature region after total 880 hour irradiation.}
\label{fig2}
\end{figure}

Figure 2 shows the temperature dependence of the in-plane resistivity in (a) $\kappa$-(ET)$_{2}$Cu[N(CN)$_{2}$]Cl (No.1) and (b) $\kappa$-(ET)$_{2}$Cu$_{2}$(CN)$_{3}$. 
Before irradiation, both insulators show an activation type behavior of the resistivity as has been reported so far.\cite{Williams,Bu} 
With increasing the irradiation dose, the resistivity decreases in the whole temperature range.  
It is noted, moreover, that a metal-like temperature dependence appears in $\kappa$-(ET)$_{2}$Cu[N(CN)$_{2}$]Cl with rather small irradiation dose.  
The resistivity decreases sublinearly with lowering temperature down to about 50 K, and then it turns to increase rapidly.  
The temperature of about 50 K corresponds also to the characteristic temperature in the non-irradiated sample at which the activation energy obtained from the Arrhenius plots of the resistivity changes as seen latter in the inset of Fig. 3, and the charge gap found in the optical measurements starts to grow.\cite{Kornelson}  
The temperature region where the metal-like behavior is observed, however, does not extend below 50 K with increasing the irradiation dose.
At last the sample irradiated for 620 hours turns to show the increase of the resistivity at low temperatures and concurrently the metal-like behavior of the resistivity becomes weak.
As shown in Fig. 2(b), the resistivity in $\kappa$-(ET)$_{2}$Cu$_{2}$(CN)$_{3}$ decreases with the irradiation as well.  
But it takes so much irradiation dose to observe the metal-like behavior of the resistivity only in the narrow temperature region above about 230 K as shown in the inset.  

The observation of the large decrease and the metal-like behavior of the resistivity in X-ray irradiated Mott insulators leads us to a possible model that carriers are introduced by the irradiation into the clean system with a high carrier mobility $\mu$.  
Here we simply consider the electronic conduction as $1 / \rho = \sigma = ne\mu$, where $n$ is the carrier number and $e$ is the electron charge. 
The resistivity in the clean band insulator at finite temperatures is described dominantly by the thermally activated carrier number with an energy gap $\Delta$ as $n \propto \exp(-{\Delta}/2k_{\rm B}T)$ and the mobility determined by electron scatterings with lattice vibrations should vary with temperature as $T^{-3/2}$.  
When additional carriers are doped into the sample, for example, by doping the intrinsic semiconductors with the impurity atoms having a different valence, the resistivity becomes decreasing with the amount of the introduced carriers. 
This mechanism of carrier doping may take place for the present X-ray irradiation effect in the organic Mott insulators and also the organic superconductor having the semiconducting behavior at high temperature region, which show the activation-type behavior of $\rho$($T$) with small $\Delta$ mentioned later.
Once carriers are introduced in addition to the thermally activated ones, the temperature dependence of the resistivity must not be so simple.  
The resistivity, however, in the sample with enough high mobility has a tendency to follow the temperature dependence of the mobility that increases with lowering temperature.  
Thus the observed metal-like behavior could be explained phenomenologically by the temperature dependence of the mobility of the introduced carriers, although the effect of the carrier doping in the Mott insulators must not be so simple.\cite{Imada}

Let us then consider a possible process for introducing carriers into the Mott insulator by X-ray irradiation.  
The X-ray irradiation to the organic material has been considered to introduce the molecular defects which are radiolysed by ionizing radiation.\cite{x-ray}  
This kind of molecular defects permanently remains, while the irradiation damage in inorganic materials in general is only due to atomic displacements which can be restored by a proper heat treatment.  
It has been also known that the average volume of the molecular defect has been in the order of one molecular volume.
For the charge transfer salts consisting of donors and accepters, it has not been clear which molecules is damaged dominantly by the irradiation.  
The irradiation effects on $T_{\rm c}$, the residual resistivity and other electronic properties should suggest that the donor molecules, i.e. ET in the present case, are damaged dominantly.
When one ET molecule in the charge transfer salt of (ET)$_{2}$$X$ is chemically damaged and converted to an alternate stuff with different valence from the pristine value of +0.5, the average carrier number should change from the value of one hole per two ET molecules.  
In the metal with the quarter-filled band, the induced damage will result in the increase of the scatterings, and the influence of the slight change of the averaged carrier number will not be so significant that the qualitative electronic property does not change.  
In the insulators, however, such kinds of carrier doping must change the transport properties drastically, which is similar to the doped semiconductors.  
Moreover the qualitative alternation over the simple carrier doping may be expected in the present dimer-Mott insulators because the strongly correlated circumstance with the half-filled band is based on the strong ET dimer structure with one hole per one dimer.  
The irradiation-induced ET defects must alter not only the carrier number but also the magnitude of $U_{\rm eff}$ on the average in the crystal because $U$ on the ET molecule, the transfer energy $t$, the Coulomb interaction between neighboring site $V$, and the screening effect by the carriers are modified around the defect sites locally. 
This point must be different from the carrier doping with impurities to a band insulator.

\begin{figure}[tb]
\begin{center}
\includegraphics[viewport=3cm 9.5cm 18cm 22.5cm,clip,width=0.9\linewidth]{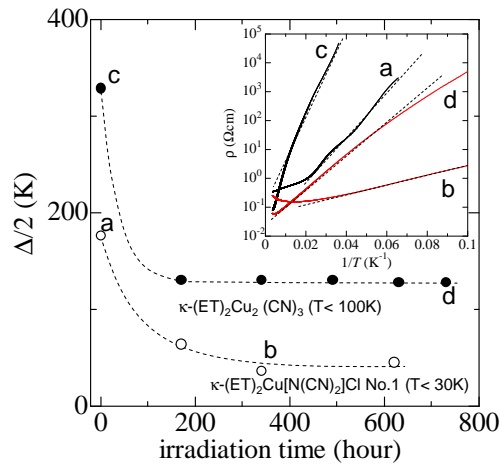}
\end{center}
\caption{(Color online) Irradiation time dependence of the activation energy in $\kappa$-(ET)$_{2}$Cu[N(CN)$_{2}$]Cl (No.1) and $\kappa$-(ET)$_{2}$Cu$_{2}$(CN)$_{3}$. Broken curves are guide for eyes.   Data points indicated as a, b, c and d correspond to the resistivity curves in the Arrhenius plots of the inset.}
\label{fig3}
\end{figure}

Figure 3 shows the variation of the activation energy at low temperatures with the irradiation time in $\kappa$-(ET)$_{2}$Cu[N(CN)$_{2}$]Cl (No.1) and $\kappa$-(ET)$_{2}$Cu$_{2}$(CN)$_{3}$.  
The activation energy $\Delta$ is obtained from the Arrhenius plot, as shown in the inset, assuming the simple form as $\rho(T) \propto \exp(\Delta/2k_{\rm B}T)$ at $T <$ 30 K and $T <$ 100 K in $\kappa$-(ET)$_{2}$Cu[N(CN)$_{2}$]Cl (No.1) and $\kappa$-(ET)$_{2}$Cu$_{2}$(CN)$_{3}$, respectively. 
The resistivity curves are not perfectly fitted to this plot as shown by the broken straight lines especially in $\kappa$-(ET)$_{2}$Cu$_{2}$(CN)$_{3}$,\cite{Tokura} but here the activation energy is deduced from the simple line fitting in order to evaluate the irradiation effect.  
It is noted that the resistivity of the irradiated $\kappa$-(ET)$_{2}$Cu[N(CN)$_{2}$]Cl shows the Arrhenius-type behavior rather than the localization one.  
This is probably because the present irradiation dose is much lower than the disorder level causing the localization.  

The activation energies decrease rapidly to 25 \% and 40 \% of the initial values in $\kappa$-(ET)$_{2}$Cu[N(CN)$_{2}$]Cl (No.1) and $\kappa$-(ET)$_{2}$Cu$_{2}$(CN)$_{3}$, respectively, and then they stay constant.
One explanation for the decrease of the activation energy is the formation of the impurity level induced by the defects in the charge gap between the Hubbard bands.
Another is a slight modification of $U_{\rm eff}$ and/or $t$ by the defect in the dimer.  
The charge gap $\simeq$ 0.1 eV has been found in $\kappa$-(ET)$_{2}$Cu[N(CN)$_{2}$]Cl below 50 K,\cite{Kornelson} which is smaller than $U_{\rm eff} \simeq$ 0.4 - 0.5 eV.\cite{Miyagawa,Kanoda2,Kino}  
The activation gap could be comparable in magnitude to the charge gap.  
The local defects in the dimers might reduce $U_{\rm eff}$ on the average in the crystal due to the screening effect by the carriers.  
Then the change of $U_{\rm eff}$ leads to the reduction of the activation energy through the charge gap. 
The preliminary magnetic susceptibility measurements in the irradiated $\kappa$-(ET)$_{2}$Cu[N(CN)$_{2}$]Cl show the increase of the N\'eel temperature $T_{\rm N}$ from 27 K before irradiation\cite{Miyagawa2} to 29 K, which results will be published elsewhere.\cite{Yoneyama2}  
The observed increase of $T_{\rm N}$ may suggest the enhancement of the exchange interaction $J$.  
This may result from the change of $U_{\rm eff}$ and/or $t$ through the simple relation $J \propto t^{2}/U_{\rm eff}$.
At present we cannot conclude, however, which scenario may be a better description for the observations.  
Spectroscopic studies like as the optical conductivity in addition to evaluate the carrier number by the Hall effect measurement as well will be important in future to understand the modification of the electronic state by the irradiation.  

The magnetic property in the irradiated $\kappa$-(ET)$_{2}$Cu$_{2}$(CN)$_{3}$ is also interesting.  
This Mott insulator has been expected to possess the spin liquid state at low temperature due to the large spin frustration in the triangular lattice.\cite{Shimizu}
The local defects will work as the symmetry-breaking site in the triangular lattice, partially breaking the spin liquid.\cite{Shimizu2}

Finally, we would like to comment on the resistivity behavior at high temperatures in the superconductors $\kappa$-(BEDT-TTF)$_{2}$$X$ with $X =$ Cu(NCS)$_{2}$ and Cu[N(CN)$_{2}$]Br.  
A variety of the temperature dependence of the resistivity has been reported so far.\cite{Strack}  
The origin of the difference has been discussed on the correlation with the sample quality from several points of view.\cite{Strack}  
This suggests that the difference of the crystal growth conditions, for example, purity of chemicals, type of the solvents, crystal growth temperature, etc., may introduce different amount of defects in the as-grown crystals. 
Such the defects in the as-grown crystals may suppress the semiconducting behavior at high temperatures as deduced from the present X-ray irradiation study and also by Analytis {\it et al.}\cite{Analytis}
The clean and defects-free sample is expected to show the clear semiconducting behavior at high temperatures. 
In order to make these points clear, it is necessary to evaluate the effect of the impurities and defects derived from different sources on the transport properties.

In conclusion, we demonstrate the carrier doping effects to the organic dimer-Mott insulators $\kappa$-(ET)$_{2}$$X$ in the process of X-ray irradiation.  
The irradiation-induced molecular defects result in the effective carrier doping into the half-filled dimer-Mott insulators.  
In addition to the introduction of carriers, the effective Coulomb energy may be modulated in the dimer-Mott insulator system. 

This work was partly supported by a Grant-in-Aid for Scientific Research (No. 16076201, 17340099 and 18654056) from MEXT and JSPS, Japan.


\begin{thebibliography}{99} %% The number "99" means that this list has more than nine items.

\bibitem{Imada} M.~Imada, A.~Fujimori and Y.~Tokura: Rev. Mod. Phys. \textbf{70} (1998) 1039.
\bibitem{Miyagawa} K.~Miyagawa, K.~Kanoda and A.~Kawamoto: Chem. Rev. \textbf{104} (2004) 5635.
\bibitem{Kanoda2} K.~Kanoda: Hyperfine Interact. \textbf{104} (1997) 235.
\bibitem{Kino} H.~Kino and H.~Fukuyama: J. Phys. Soc. Jpn. \textbf{65} (1996) 2158.
\bibitem{Ito} H.~Ito, T.~Ishiguro, M.~Kubota and G.~Saito: J. Phys. Soc. Jpn. \textbf{65} (1996) 2987.
\bibitem{Lefebvre} S.~Lefebvre, P.~Wzietek, S.~Brown, C.~Bourbonnais, D.~J\'erome, C.~M\'ezi\`ere, M.~Fourmigu\'e and P.~Batail: Phys. Rev. Lett. \textbf{85} (2000) 5420.  
\bibitem{Yoneyama1} N.~Yoneyama, T.~Sasaki and N.~Kobayashi: J. Phys. Soc. Jpn. \textbf{73} (2004) 1434.
\bibitem{FET} M.~E.~Gershenson, V.~Podzorov and A.~F.~Morpurgo: Rev. Mod. Phys. \textbf{78} (2006) 973.
\bibitem{Mori} T.~Mori: Chem. Rev. \textbf{104} (2004) 4947. 
\bibitem{x-ray} L.~Zuppiroli: Radiation Effects \textbf{62} (1982) 53.
\bibitem{TMTSF} M.~-Y.~Choi, P.~M.~Chaikin, S.~Z.~Huang, P.~Haen, E.~M.~Engler and R.~L.~Greene: Phys. Rev. B \textbf{25} (1982) 6208.
\bibitem{BEDT} S.~D.~Babi\'c, N.~Bi\v skup, S.~Tomi\'c and D.~Scweitzer: Phys. Rev. B \textbf{46} (1992) 11765.
\bibitem{Analytis} J.~G.~Analytis, A.~Ardavan, S.~J.~Blundell, R.~L.~Owen, E.~F.~Garman, C.~Jeynes and B.~J.~Powell: Phys. Rev. Lett. \textbf{96} (2006) 177002.
\bibitem{Strack} Ch.~Strack, C.~Akinci, V.~Pashchenko, B.~Wolf, E.~Uhrig, W.~Assmus, M.~Lang, J.~Schreuer, L.~Wiehl, J.~A.~Schlueter, J.~Wosnitza, D.~Schweitzer, J.~M\"uller and J.~Wykhoff: Phys. Rev. B \textbf{72} (2005) 054511, and references therein.
\bibitem{Mueller} J.~M\"uller, M.~Lang, F.~Steglich, J.~A.~Schlueter, A.~M.~Kini and T.~Sasaki: Phys. Rev. B \textbf{65} (2002) 144521.
\bibitem{Powell2} B.~J.~Powell and R.~H.~Mckenzie: Phys. Rev. B \textbf{69} (2004) 024519.
\bibitem{BETS} A.~Kobayashi, T.~Udagawa, H.~Tomita, T.~Naito and H.~Kobayashi: Chem. Lett. \textbf{1993} (1993) 2179.
\bibitem{Williams} J.~M.~Williams, A.~M.~Kini, H.~H.~Wang, K.~D.~Carlson, U.~Geiser, L.~K.~Montgomery, G.~J.~Pyrka, D.~M.~Watkins, J.~M.~Kommers, S.~J.~Boryschuk, A.~V.~S.~Crouch, W.~K.~Kwok, J.~E.~Schirber, D.~L.~Overmyer, D.~Jung and M.~-H. Whangbo: Inorg. Chem. \textbf{29} (1990) 3272.
\bibitem{Bu} X.~Bu, A.~Frost-Jensen, R.~Allendoerfer, P.~Coppens, B.~Lederle and M.~J.~Naughton: Solid State Commun. \textbf{79} (1991) 1053.
\bibitem{Kornelson} K.~Kornelson, J.~E.~Eldridge, H.~H.~Wang, H.~A.~Charlier and J.~M.~Williams: Solid State Commun. \textbf{81} (1992) 343. 
\bibitem{Tokura} I.~K\'ezsm\'arki, Y.~Shimizu, G.~Mih\'aly, Y.~Tokura, K.~Kanoda and G.~Saito: Phys. Rev. B \textbf{74} (2006) 201101(R).
\bibitem{Miyagawa2} K.~Miyagawa, A.~Kawamoto, Y.~Nakazawa and K.~Kanoda: Phys. Rev. Lett. \textbf{75} (1995) 1174.
\bibitem{Yoneyama2} N.~Yoneyama, T.~Sasaki and N.~Kobayashi: unpublished.
\bibitem{Shimizu} Y.~Shimizu, K.~Miyagawa, K.~Kanoda, M.~Maesato and G.~Saito: Phys. Rev. Lett \textbf{91} (2003) 107001.
\bibitem{Shimizu2} Y.~Shimizu, K.~Miyagawa, K.~Kanoda, M.~Maesato and G.~Saito: Phys. Rev. B \textbf{73} (2006) 140407(R).
\end{thebibliography}
\end{document}